\documentstyle{l-aa}

\begin{document}

   \thesaurus{05         % A&A Section 5: Stellar clusters and associations
              (08.12.1;           % Stars: late-type,
               08.12.2;           % Stars: low-mass, brown dwarfs,
               08.12.3;           % Stars: luminosity function, mass function,
               10.15.2 Pleiades)}  % open clusters and associations:individual:
%%Pleiades
%
   \title{Brown Dwarfs in the Pleiades\thanks{Based on observations made with
the Nordic Optical Telescope, La Palma.}}

   \subtitle{A deep $IJK$ survey}

   \author{L. Festin
%          \inst{1}
          }

   \offprints{L. Festin}

   \institute{Astronomical Observatory,
	      Box 515,
	      S-751 20 UPPSALA,
              Internet: leif@astro.uu.se
             }

   \date{Received            ; accepted           }

   \maketitle

   \begin{abstract}
An area large enough (180 $\mbox{arcmin}^{2}$) to put constraints on a possible
low mass brown dwarf population in the Pleiades has been surveyed to very faint
magnitudes in $I$, $J$ and $K$. The completeness limit, I=21.6, corresponds to
a mass of 0.01 $\mbox{M}_{\odot}$ for a cluster age of 70 Myr and 0.035
$\mbox{M}_{\odot}$ for 120 Myr. The result is consistent with previous
investigations at higher masses that the brown dwarf initial mass function is a
$m^{-1}$, or even less steep, power law. Thus low mass brown dwarfs cannot
contribute significantly to the Pleiades' mass. One new possible Pleiades
member was found, mass $\sim0.08$ $\mbox{M}_{\odot}$ (age 120Myr).
   \end{abstract}

%
%________________________________________________________________

\section{Introduction}
Brown dwarfs (BDs) are stellar-like objects. The only difference from ordinary
stars is that the mass is too low to bring up the central temperature to the
level of stable hydrogen burning, thus the BD luminosity decreases with time.
As an example, from 70 Myr to 10 Gyr, a 0.08 $\mbox{M}_{\odot}$ object at the
hydrogen burning limit would decrease a factor 15 in luminosity, while a 0.06
$\mbox{M}_{\odot}$ would go a factor 700 (Burrows et al. 1993).
The ideal target for a BD search thus is a fairly young, nearby and rich star
cluster. The Pleiades is the obvious choice in the northern hemisphere, being
at $\sim 125$ pc and 70-120 Myr old. Several recent authors have proposed an
age above 100 Myr. In this paper 120 Myr is adopted. The mean distance modulus
of the cluster used is 5.53 (see Basri et al. 1996 for a discussion).

Whether BDs could be a significant part of the local dark matter is a subject
of controversy. Several recent photometric surveys have found a significant
drop in the luminosity function from $M_{V}=12$ to $M_{V}=14$, leading to a
turnover in the initial mass function (IMF, $dN_{\rm stars} = const*m^{-n}*dm$
($m$ = mass;  $n$ = IMF-index)) at $\sim$ 0.3 $\mbox{M}_{\odot}$ (see e.g.
Gould et al. 1996; Tinney 1993) or a continued rise towards the hydrogen
burning limit (see e.g. Kirkpatrick et al. 1994) depending on the choice of
mass-luminosity relation. Kroupa (1995) showed that the difference between the
nearby stellar luminosity function (LF), measured by parallax and the more
distant LF, measured by photometry alone, can be explained by undetected binary
companions in the distant sample.
{}From a recently derived mass-luminosity relation (Chabrier et al. 1996) and
the well-known photometric LF (see e.g. Gould et al. 1996), Mera et al. (1996)
concluded that the IMF for low mass stars continues to rise to the hydrogen
burning limit. However the number of known field stars at the low mass end of
the main sequence is small. To clearify this item it is necessary to discover
more low mass stars and BDs, preferrably at a known distance and age.

In \S 2, observations, reductions and a short discussion on photometry and
completeness limits is given. In \S 3, the extraction of Pleiades members is
described. \S 4 discusses contamination and overall observing strategy. In \S 5
the implications of this paper on the IMF-index is discussed and compared to
other authors.

\section{Observations and Reductions}
All observations for this program were obtained at the 2.56 m Nordic Optical
Telescope (NOT), La Palma. The observed area covers 180 $\mbox{arcmin}^{2}$ in
$IJK$ and is centred at RA $3^{\mbox{h}}48^{\mbox{m}}3.6^{\mbox{s}}$, Dec
$23^{\degr}44'13.1"$ (J2000.0).

The $I$ data was taken with BROCAM1 (TEK1K), operated in cassegrain focus with
0.176"/pixel and 3'x3' field. For future proper motion measurements, the
astrometry errors due to nonuninform pixels and possible effects of rotator
position were reduced by using four different position angles for each I-field
(0, 90, 180 and 270 degrees) exposing 5 min at each position. Debiassing and
flatfield corrections were done in a standard fashion within IRAF (Image
Reuction and Analysis Facility) \footnote{IRAF is distributed by National
Optical Observatories (NOAO), which is operated by the Association of
Universities for Research in Astronomy, Inc., under contract with the National
Science Foundation.}. Median seeing was $\sim 0.6$", varying from 0.46" to
0.80".

The $JK$ observations were done with the ARNICA NICMOS3 (256x256) array, which
was made available at NOT via a collaboration with Arcetri Astrophysical
Observatory, Florence. The last third of 6 nights in Aug-Sep95 was used.
Unfortunately the pixel size (0.55") was not very well matched to the seeing
($<0.8"$) and all images were undersampled. Another problem was that focus
changed across the field, which due to astigmatism in the NOT optics led to
stars of slightly different elongation across the field. Apparently this was
caused by the chip not being perpendicular to the optical axis which could be
seen as a slight change in pixel scale across the field.

The $J$ data could be treated the same way as $I$, but high and rapidly varying
background in $K$ necessitated a different approach. The sky for each $K$-image
was defined as the median of the four images nearest in time and subtracted
from the image, which then was flatfielded by a differential flat.

\subsection{classification of objects}
A major problem in this kind of survey is to distinguish stars from
distant galaxies at faint magnitudes. The whole area was visually inspected and
all objects were classified as stars, galaxies, binaries or unclassifiable (too
faint for a meaningful classification). 1411 objects out of 3800 were
classified as stars. Since the seeing conditions were excellent it is estimated
that the discrimination between stars and galaxies is reliable to $I \sim 22$.

\subsection{photometry}
Since point spread functions (PSFs) were undersampled in $JK$ it was not
suitable to use ordinary PSF fitting. For single aperture photometry it is
necessary to always keep the same aperture radius. However for the $JK$ data
centering errors and focus shifts are non-negligible, thus in order to maximize
the signal to noise ratio it was desirable to change aperture radius from one
object to another. A new, more robust, method (used also in $I$) was developed,
brief outlines:

   \begin{enumerate}
      \item All objects were measured in a series of 80 apertures, with radii
            ranging from 0.05" to 4" in steps of 0.05".
      \item From the "best" stars in each field a growth curve (Stetson 1990)
            was constructed, i.e. magnitude shift as function of aperture
            radius.
      \item Each objects' curve was subtracted by that frames' growth curve.
            Ideally this curve is completely flat, but due to centering errors,
            errors in the sky level etc... that is generally not the case.
            However, there is normally an almost flat part of the curve at
            $\sim1-2$ FWHM. The magnitude difference was
	    taken as the mean level of that part. The advantage with this
	    method is that the right mean level will be found even if
	    objects are differently focussed or elongated as compared to the
	    reference growth curve.
   \end{enumerate}

Transformation of $JK$ instrument magnitudes to the CIT system was done by
observations of IR standard stars and transformation equations provided by the
ARNICA team (Hunt et al. subm.). $I$ instrument magnitudes were transformed to
the Kron-Cousins system via observations of Landolt (1992) standard stars.
Typical internal errors are 0.1 mag ($I=21.9$; $J=19.3$; $K=16.9$) and 0.02 mag
($I=19.8$; $J=16.7$; $K=14.9$). The zeropoint errors are $\sim$0.05 mag in J
and $\sim$0.08 mag in K. In I they are estimated to be less than 0.02 mag.
Zeropoint errors are not included in Table 1 and Table 3.

All objects recognized as binaries or stars very close to a galaxy in $I$ were,
as a check, also measured by PSF-fitting. For a few objects that were resolved
in $I$, but not in $JK$, the whole system was measured as one object in all
bands, thus getting the binary system correct in the colour magnitude diagrams.

\subsection{completeness limits}
In this kind of surveys it is important to know the magnitude limit to which
the survey is complete.
The number of detected stars as a function of magnitude increases exponentially
and gives a straight line in a log($N_{stars}$) vs mag plot as long as the
survey is complete. The point at which the curve turns away from the straight
line should be regarded as the completeness limit. Thus $I=21.6$, $J=18.9$ and
$K=16.7$ are the completeness limits of this survey (magnitude error $\sim$
0.08 mag). The 50 \% limit is $\sim$ 0.7 mag fainter.

\section{Analysis}
\subsection{Extraction technique}

\begin{figure}
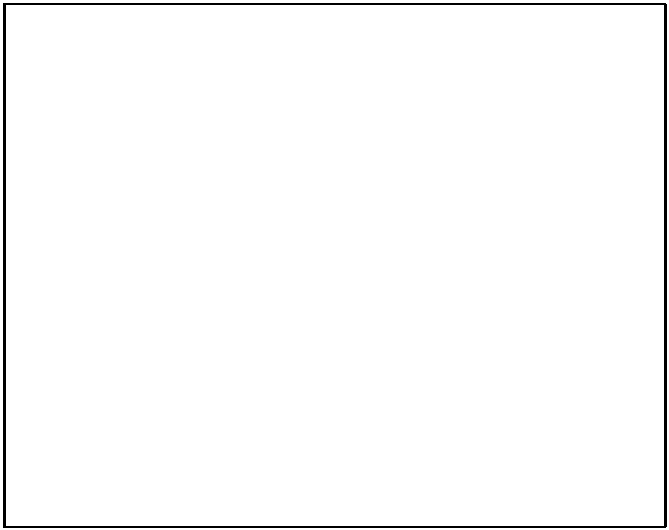

\picplace{6.95 cm}
\caption{Colour-magnitude diagram for all stars detected in both $I$ and $J$.
NOT1 is a Pleiades candidate and other NOT stars are probable background
M-dwarfs. Ticks ($\mbox{M}_{\odot}$): 0.035, 0.045, 0.055, 0.060, 0.070, 0.075,
0.078, 0.080, 0.090, 0.100, 0.125, 0.150, 0.200 (Allard) and 0.045, 0.060,
0.070, 0.075, 0.080, 0.090, 0.100, 0.125, 0.150, 0.200, 0.220 (Brett,Plez). The
two field stars are BRI0021-0214 ($>$M9.5V) and RG0050-2722 (M8V), reduced in
the same way as the program fields. The models and the two field stars have
been shifted to the distance and general reddening of the Pleiades.}
\end{figure}

Several different kinds of errors define the zone of potential Pleiades members
in the colour magnitude diagram. The maximum error in the absolute magnitude of
a single Pleiades member is not likely to exceed 0.2 mag because of uncertainty
of its distance. The photometric error for $I < 21.0$ is less than 0.1 mag. For
binaries, the worst case is considered. Two identical companions get 0.75 mag
brighter than a single star of the same colour. Therefore the bright edge is
raised by 0.75 mag. The age of the Pleiades is also a source of error. Basri et
al. (1996) argue that the age is $\sim$ 115 Myr while the "canonical" age
deduced from the upper main sequence turn-off is 75 Myr. To account for the
worst possible case, the 70 Myr sequence -1.05 mag defines the bright limit,
and the 120 Myr model +0.30 mag defines the faint limit. An extension of the
model colours in Chabrier et al. (1996) based on the Brett (1995) models was
kindly provided by Plez (1996). Next generation models from Allard et al.
(1996) were also included. Both models show excellent agreement with the data
(Fig. 1 \& 2).

Note the clear gap between the background field stars and the Pleiades sequence
and that all previously known members within the field were detected with this
technique.

\begin{figure}
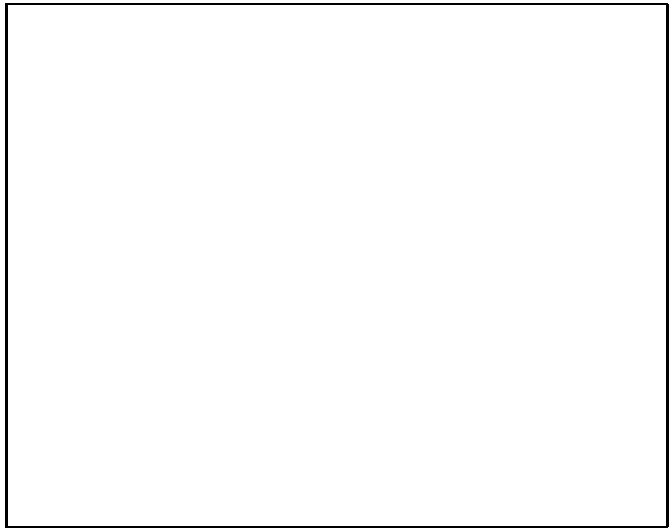

\picplace{6.95 cm}
\caption{All stars detected in both $I$ and $K$. Symols are the same as in Fig.
1.}
\end{figure}

\subsection{Extracted objects}
As shown in Fig. 1 \& 2, 6 objects in the sample ($I>15$) can be regarded as
potential cluster members (Table 1). Of these, 5 were previously known as
proper motion members (Hambley et al. 1993; Rebolo et al. 1995). The new
candidate (NOT1) is very close to a much brighter star, discovered only as a
consequence of the excellent seeing conditions during the observations. A
finding chart and coordinates of NOT1 are given in Fig. 4. Its colours and
magnitude indicate a binary of equally sized components, each of $\sim 0.08
\mbox{M}_{\odot}$, slightly heavier than PPL15 (Basri et al. 1996). It is,
however, also consistent with being a foreground M-dwarf (see Sect. 4).

\begin{table}
\caption[]{Photometry of potential Pleiades members. Magnitude errors are
internal.\\
}
\begin{flushleft}
\begin{tabular}{llll}
\hline
id & $I_{\rm KC}$  & $J_{\rm CIT}$ & $K_{\rm CIT}$  \\
\hline
 HHJ424 & saturated     & saturated     & 10.99 (0.01) \\
 HHJ336 & saturated     & 12.74 (0.01)  & 11.99 (0.01) \\
 HHJ288 & saturated     & 13.25 (0.01)  & 12.36 (0.01) \\
 HHJ152 & 15.14 (0.01)  & 13.55 (0.01)  & 12.82 (0.01) \\
 HHJ188 & 15.26 (0.01)  & 13.78 (0.01)  & 12.95 (0.01) \\
 HHJ156 & 15.26 (0.01)  & 13.77 (0.01)  & 12.90 (0.01) \\
 HHJ122 & 15.55 (0.01)  & 13.90 (0.01)  & 13.13 (0.01) \\
 NOT1   & 16.91 (0.01)  & 14.61 (0.01)  & 13.58 (0.01) \\
 PPL15  & 17.91 (0.01)  & 15.43 (0.01)  & 14.48 (0.02) \\
 JS3    & 17.72 (0.05)  & 16.23 (0.01)  & 15.25 (0.02) \\
 JS5    & 17.06 (0.05)  & 15.60 (0.01)  & 14.68 (0.02) \\
\hline
\end{tabular}
\end{flushleft}
\end{table}
The last two stars in Table 1 were previously regarded as Pleiades candidates
from $RI$ photometry (Jameson \& Skillen 1989), but can be judged as clear
nonmembers from Fig. 1, which is the same result as Zapatero Osorio et al.
(1996) got from their $RI$ survey.

\subsection{Comparison with other works}
Several papers have recently been published on the Pleiades. The proper motion
survey by Hambley et al. (1993), using POSS and UKSTO Schmidt plates, covered
most of the cluster, complete to $I\sim 16.5$, corresponding to a star of 0.10
$\mbox{M}_{\odot}$, slightly above the hydrogen burning limit. Their limiting
magnitude was $\sim 1$ mag fainter and they reached objects of $\sim$ 0.08
$\mbox{M}_{\odot}$, right at the BD limit.

Jameson \& Skillen (1989) imaged 175 $\mbox{arcmin}^{2}$ in $RI$. Some of their
fields overlap with ours and after checking their original data, we concluded
that their completeness limit is $I\sim20.7$ and $R\sim23.1$. Their R limit
corresponds to $I\sim 20.3$ on the Pleiades sequence. They found 9 BD
candidates, all of which later have been shown to fall significantly below the
Pleiades sequence in colour magnitude diagrams (Zapatero Osorio et al. 1996).
Rebolo et al. (1995) announced the discovery of a BD (Teide1) in the Pleiades
as a result of a CCD survey, 175 $\mbox{arcmin}^{2}$. This survey was later
extended to 578 $\mbox{arcmin}^{2}$, complete to $I\sim19.5$, and another
candidate (Calar3) was found (Zapatero Osorio et al. 1996). Both candidates
have passed all membership tests hitherto, including the lithium test (Rebolo
et al . 1996) and can be regarded as genuine BDs.
Stauffer et al. (1994) surveyed $\sim$ 1500 $\mbox{arcmin}^{2}$. Their survey
is complete to ${V} \sim 22$ corresponding to ${I} \sim 18$ in the Pleiades
domain of the $V$ vs $V-I$ diagram. They found 15 very red Pleiades candidates,
among them PPL15, which for a cluster age of 75 Myr would be of mass $\sim0.06$
$\mbox{M}_{\odot}$. However, Basri et al. (1996) observed PPL15 at Keck and
suggested, based on the lithium line strength, that its age is around 115 Myr,
thus increasing the mass from 0.065 $\mbox{M}_{\odot}$ to 0.077
$\mbox{M}_{\odot}$. Basri also concludes, based on the non-detection of lithium
in the BD candidate HHJ3 (Hambley et al. (1993)), that it cannot be younger
than 110 Myr. Thus the Pleiades age is confined approximatively between 110 Myr
and 125 Myr.
Simons \& Becklin (1992) used $I$ vs $I-K$ diagrams and compared fields within
the cluster to reference fields outside the cluster. They found an excess of
$22 \pm 10$ BD candidates in 200 $\mbox{arcmin}^{2}$ in the cluster field.
Williams et al. (1996) found 8 stars in the magnitude interval $16<I<19$, using
$V$ and $K$ instead of $IJK$, a result consistent with ours.
Steele et al. (1993) present $\sim 35$ candidates with $IJK$ photometry in the
same magnitude range, a subsample from Hambley et al. (1993), complete to
$I\sim16.5$ and therefore not included in Table 2.
All surveys except Simons \& Becklin (1992) agree reasonably well within error
bars.

\begin{table}
\caption[]{The number of Pleiades candidates in the magnitude interval
$16<I<19$ from five different surveys and as expected from four different
IMF-indices. The rightmost column is scaled to 500 $\mbox{arcmin}^{2}$.
Normalization of the IMF was deduced from proper motion members in
$15.0<I<16.4$ (Hambley et al. 1993). Statistical errors are within the
parenthesis.}
\begin{flushleft}
\begin{tabular}{llll}
\hline
origin & $\mbox{arcmin}^{2}$ & candidates & scaled \\
\hline
this paper                   &  180 &  2  (2) &  5.6 (4) \\
Jameson \& Skillen 1989      &  175 &  0  (-) &  0   (-) \\
Simons \& Becklin 1992       &  200 & 22 (10) & 55  (25) \\
Stauffer et al. 1994         & 1500 &  9  (3) &  3.0 (1) \\
Williams et al 1996          &  400 &  8  (3) & 10   (3) \\
Zapatero O et al. 1996       &  578 &  5  (2) &  4.3 (2) \\
IMF-index, $n=-2.8$          &  --- &  ------ &   37-56  \\
              -2.0           &  --- &  ------ &   16-22  \\
              -1.0           &  --- &  ------ &     6-7  \\
              -0.0           &  --- &  ------ &     2-3  \\
\hline
\end{tabular}
\end{flushleft}
\end{table}

\section{Discussion}
Photometry alone can not extract Pleiades members to a very high confidence
level, but is the most time-efficient tool for finding potential members, whose
proper motion later can be measured. In this survey it is shown that the $I$ vs
$I-J$ diagram clearly sorts out the Pleiades sequence from the background.
As an example of efficiency, consider the completeness limit of this survey
($I=21.6$), reached in 20 minutes integration at seeing 0.6". The same BD in
$V$ ($V\sim26$) would demand $\sim 8$ hours and in $R$ ($R\sim24$) $\sim 1$
hour. In $J$ it takes only $\sim$ 4 minutes. Thus even though CCDs have larger
surface area, they cannot compete with IR-arrays for objects as red as presumed
BDs. In case of $K$, the needed integration time is $\sim  8$ minutes. Adding
more complicated reductions to that and the fact that $I$ and $J$ is enough to
extract the Pleiades sequence there is no need for $K$ data in the first step
of a survey like this.

Possible contamination of the Pleiades candidates could come from reddened
background giants, M-dwarfs and unresolved galaxies. The number of
contaminating background giants in the surveyed area was estimated from the
model by Bahcall \& Soneira (1984) to be $\sim$ 0.03, thus being negligible.
The interstellar extinction adopted was $A_{I}=0.57$ mag/kpc ($E_{I-K}\sim0.43$
mag/kpc) (Lucke 1978; Cardelli et al. 1989) out to 2 kpc, then decreasing
exponentially with the same scale-factor as for the giants, which is expected
to be an upper limit of the possible extinction, and thus gives an upper limit
of the number of contaminating giants.
Due to the excellent seeing during the observing run it is believed that the
contamination by galaxies on the stellar sample is negligible for $I \la 22$.

\begin{table}
\caption[]{M-dwarf candidates, including the Pleiades candidate NOT1. Magnitude
errors are internal. Distance errors are of the order of 10\%.}
\begin{flushleft}
\begin{tabular}{lllll}
\hline
id & $I_{\rm KC}$ & $I-J_{\rm CIT}$ & $I-K_{\rm CIT}$ & d(pc)\\
\hline
 NOT1 & 16.90 (0.01) & 2.29 (0.01) & 3.15 (0.01) &  70 \\
 NOT2 & 22.29 (0.11) & 3.08 (0.23) & 4.91 (0.24) & 330 \\
 NOT3 & 21.79 (0.13) & 3.43 (0.14) & 5.08 (0.15) & 250 \\
\hline
\end{tabular}
\end{flushleft}
\end{table}
The LF from Gould et al. (1996) was used to estimate the number of foreground
M-dwarfs for $I-J>2.2$ to $<0.2$. Thus it is not likely, although possible,
that NOT1 is a foreground M-dwarf. Since late M-dwarfs are presumed to be rare,
it is interesting that a few candidates turn up in this survey, see Table 3.
Both NOT2 and NOT3 appear quite close to the Pleiades zone in Fig. 2. Could
they be low mass Pleiades BDs? NOT3 was in a part of the field that was also
covered by Jameson \& Skillen (1989). It was found in their original data,
however far too faint for astrometry of sufficiently high accuracy for a proper
motion measurement. NOT2 and NOT3 will though be observed in future runs.\\
Towards this part of the Pleiades the total reddening is $E_{B-V}\sim 0.04$ mag
(Stauffer \& Hartmann 1987), which corresponds to $E_{I-J}\sim 0.02$ mag
(Cardelli et al. 1989). Thus if these stars really are M-dwarfs, $E_{I-J}$ is
less than 0.1 mag even for the most distant candidate and does not affect
classification significantly.

\section{Conclusions}
Our goal was not to cover as large an area as possible, but to reach very faint
magnitudes in order to investigate the presence of a population of low mass BDs
in the Pleiades. This survey is complete to $I = 21.6$, corresponding to 0.035
$\mbox{M}_{\odot}$ (120 Myr) or 0.01 $\mbox{M}_{\odot}$ (70 Myr). In Fig. 3,
LFs from recent surveys are compared to LFs deduced from 4 different
IMF-indices.
\begin{figure}
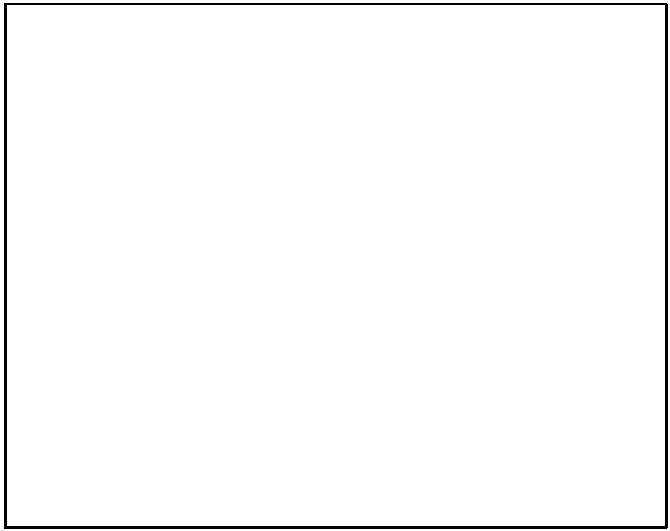

\picplace{6.95 cm}
\caption{Comparing LF predicted for four different IMF-indices (bars) and five
recent surveys (curves). All LFs were normalized to 10 counts in $16<I<17)$.
Indicated are also approximate completeness limits for the surveys. Jameson \&
Skillen (1989) (J89) is only indicated for completeness limit, since all of
their candidates fall 1.5 - 2 mag below the Pleiades sequence.}
\end{figure}
It is clear from previous surveys that an IMF-index close to or even above 0 is
favored at least down to $I \sim 18$. This survey, complete to $I=21.6$ in the
Pleiades domain, one magnitude fainter than any previous survey does not give
one single object below $I=18$.
Thus $n=-2.8$ (Simons \& Becklin 1992) can be rejected with 99.8\% confidence,
$n=-2$ with 96.9\%. Corresponding figures for $n=-1$ is 75.1\% and for $n=0$
45.0\%. The final conclusion is that if low mass BDs ($M < 0.05$
$\mbox{M}_{\odot}$) exist in the Pleiades, they follow an IMF with index less
steep than $n=-1$, and cannot comprise a very large part of the cluster's mass.

\begin{figure}
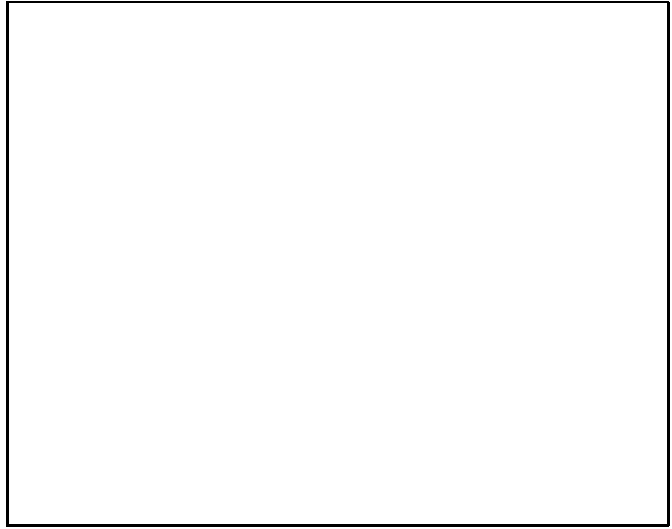

\picplace{6.95 cm}
\caption{Finding chart for the Pleiades candidate NOT1 (RA
$3^{\mbox{h}}48^{\mbox{m}}3.61^{\mbox{s}}$, Dec $23^{\degr}44'13.1"$ (J2000.0,
Epoch 1995.9), errors $\sim 1"$ in both coordinates). The size of the field is
$\sim$ 2'x2'. North is down. East is right.}
\end{figure}

\begin{acknowledgements}
      This work was supported by the Nordic Optical Telescpe Scientific
      Association and the Swedish Natural Science Research Council (NFR). This
research has made use of the La Palma archive.
\end{acknowledgements}


\begin{thebibliography}{}
        \bibitem[1996]{alla}Allard F., Alexander D.R., Hauschildt P.H.,
Schweitzer A. 1996, ApJ submitted
        \bibitem[1984]{bach}Bahcall J.N., Soneira R.M. 1984, ApJS 55, 67
        \bibitem[1996]{basr}Basri G., Marcy G.W., Graham J.R. 1996, ApJ 458,
600
        \bibitem[1995]{bret}Brett J.M. 1995, A\&A 295, 736
	\bibitem[1993]{burr}Burrows A., Hubbard W.B., Saumon D., Lunine J.I. 1993, ApJ
406, 158
	\bibitem[1989]{card}Cardelli J.A., Clayton G.C., Mathis J.S. 1989, ApJ 345,
245
	\bibitem[1996]{chab}Chabrier G., Baraffe I., Plez B. 1996, ApJ 459, L91
        \bibitem[1996]{goul}Gould A., Bahcall J.N., Flynn C. 1996, ApJ 465, 759
	\bibitem[1993]{hamb}Hambley N.C., Hawkins M.R.S., Jameson R.F. 1993, A\&AS
100, 607
	\bibitem[1995]{hunt}Hunt L.K., Migliorini S., Testi L., et al. 1995, AJ
submitted
	\bibitem[1989]{jame}Jameson R.F., Skillen I. 1989, MNRAS 239, 247
	\bibitem[1995]{krou}Kroupa P. 1995, ApJ 453, 358
	\bibitem[1994]{kirk}Kirkpatrick J.D., McGraw J.T., Hess T.R., Liebert J.,
McCarthy Jr, D.W. 1994, ApJS 94, 749
	\bibitem[1992]{land}Landolt A.U. 1992, AJ 104, 340
	\bibitem[1992]{legg}Leggett S.K. 1992, ApJS 82, 351
	\bibitem[1978]{luck}Lucke P.B. 1978, A\&A 64, 365
	\bibitem[1996]{mera}Mera D., Chabrier G., Baraffe I. 1996, ApJ 459, L87
	\bibitem[1996]{plez}Plez B. 1996, private communication
        \bibitem[1995]{reb1}Rebolo R., Zapatero Osorio M.R., Martin E.L. 1995,
NATURE 377, 129
	\bibitem[1996]{reb2}Rebolo R., Martin E.L., Basri G., Marcy G.W., Zapatero
Osorio M.R. 1996, ApJ 469, L53
	\bibitem[1992]{simo}Simons D.A., Becklin E.E. 1992, ApJ 390, 431
	\bibitem[1987]{sta1}Stauffer J.R., Hartmann L. 1987, ApJ 318, 337
        \bibitem[1994]{sta2}Stauffer J.R., Hamilton D., Probst R.G. 1994, AJ
108, 155
        \bibitem[1993]{stee}Steele I.A., Jameson R.F., Hambley N.C. 1993, MNRAS
263, 647
	\bibitem[1990]{stet}Stetson P.B., 1990, PASP 102, 932
	\bibitem[1993]{tinn}Tinney C.G. 1993, ApJ 414, 279
	\bibitem[1996]{will}Williams D.M., Boyle R.P., Morgan W.T., et al. 1996, ApJ
464, 238
	\bibitem[1996]{zapa}Zapatero Osorio M.R., Rebolo R., Martin E.L. 1996, A\&A in
press
\end{thebibliography}
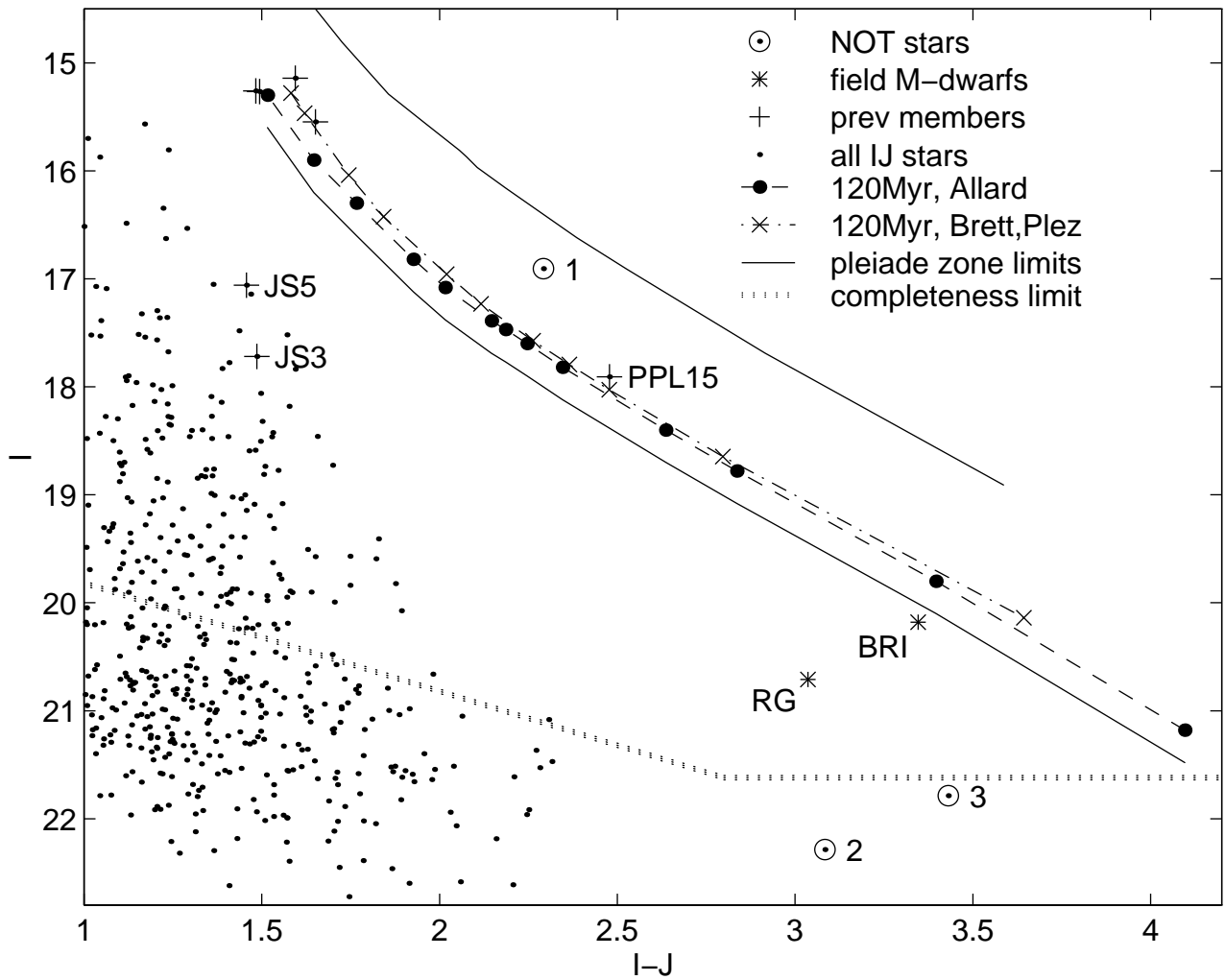
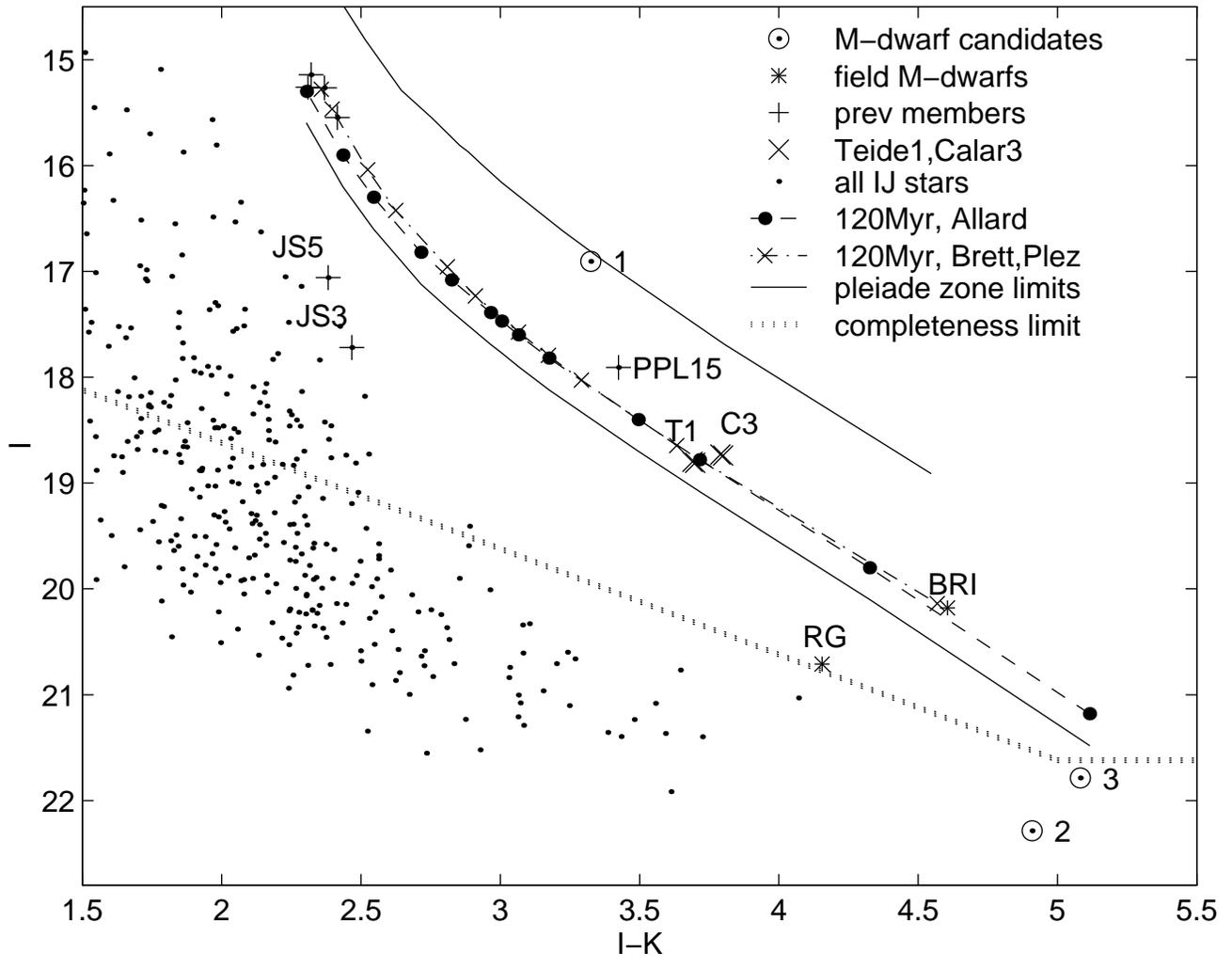
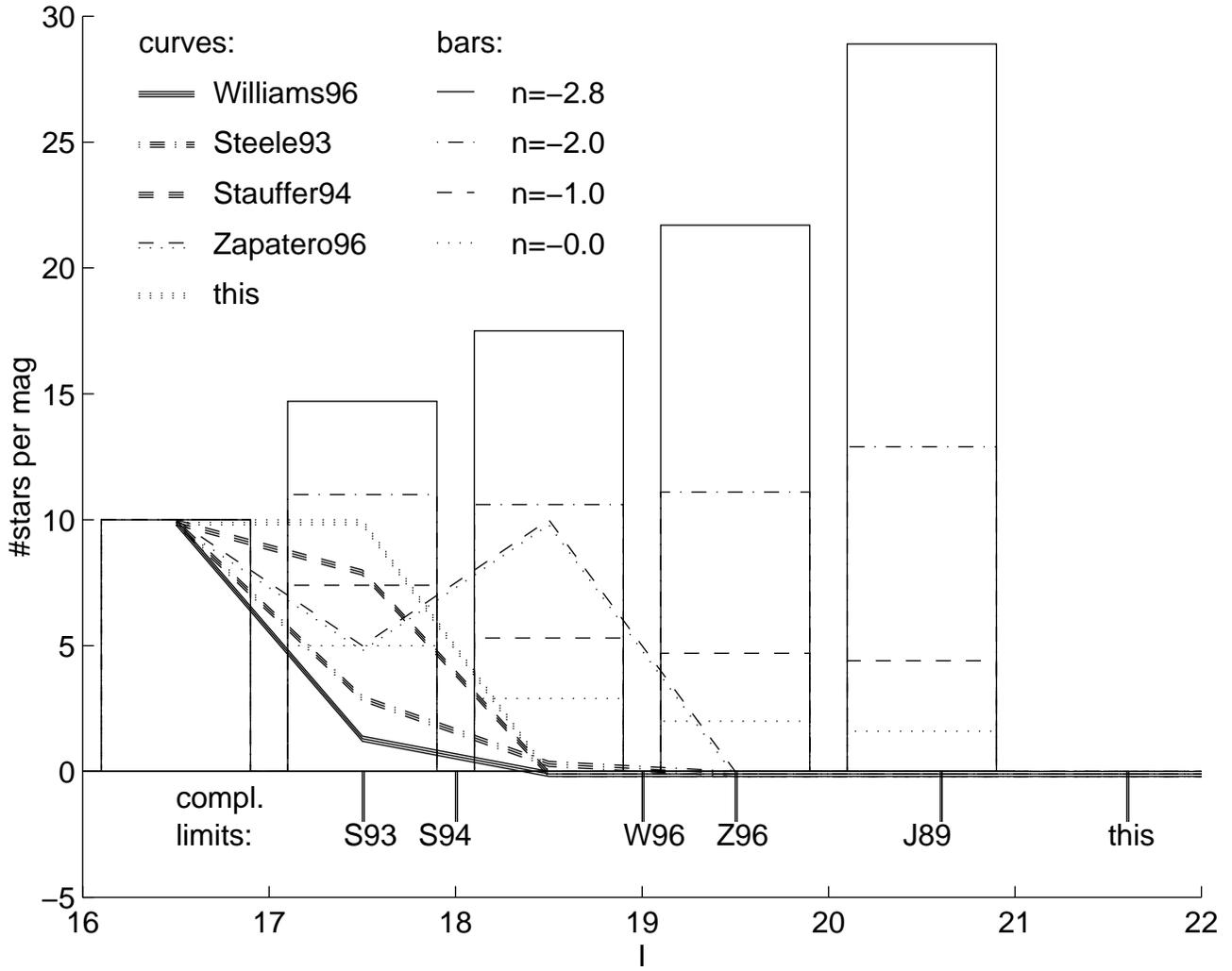
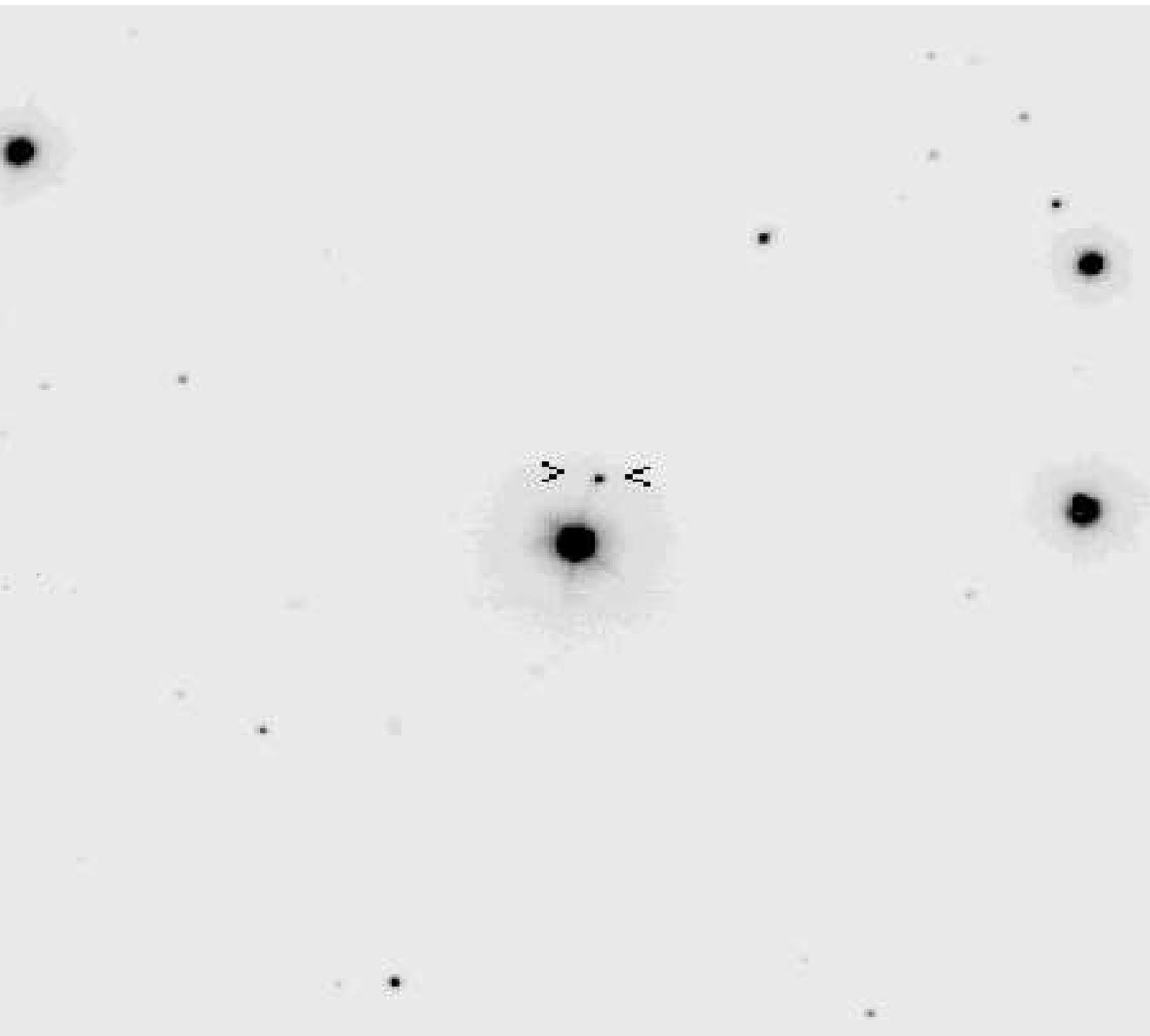
\end{document}